# Ultrafast Preparation and Detection of Ring Currents in Single Atoms

Sebastian Eckart[1*], Maksim Kunitski[1], Martin Richter[1], Alexander Hartung[1], Jonas Rist[1], Florian Trinter[1], Kilian Fehre[1], Nikolai Schlott[1], Kevin Henrichs[1], Lothar Ph. H. Schmidt[1], Till Jahnke[1], Markus Schöffler[1], Kunlong Liu[2], Ingo Barth[2], Jivesh Kaushal[3], Felipe Morales[3], Misha Ivanov[3], Olga Smirnova[3], and Reinhard Dörner[1*]

[1]Institut für Kernphysik, Goethe-Universität Frankfurt am Main, Max-von-Laue-Straße 1, 60438 Frankfurt am Main, Germany.

[2]Max Planck Institute of Microstructure Physics, Weinberg 2, 06120 Halle (Saale), Germany.

[3]Max Born Institute, Max-Born-Str. 2A, 12489, Berlin, Germany.

*Corresponding authors: e-mail: (S.E.) eckart@atom.uni-frankfurt.de; (R.D.) doerner@atom.uni-frankfurt.de

Quantum particles can penetrate potential barriers by tunneling[1]. If that barrier is rotating, the tunneling process is modified[2,3]. This is typical for electrons in atoms, molecules or solids exposed to strong circularly polarized laser pulses[4,5]. Here we measure how the transmission probability through a rotating tunnel depends on the sign of the magnetic quantum number $m$ of the electron and thus on the initial sense of rotation of its quantum phase. We further show that the electron keeps part of that rotary motion on its way through the tunnel by measuring $m$-dependent modification of the electron emission pattern. These findings are relevant for attosecond metrology as well as for interpretation of strong field electron emission from atoms and molecules[6-13] and directly demonstrates the creation of ring currents in bound states of ions with attosecond precision. In solids, this could open a way to inducing and controlling ring-current related topological phenomena[14].

Within the Bohr model, the electron travels around the nucleus on circular orbits possessing quantized orbital angular momentum associated with a ring current. In quantum mechanics this motion is reflected by the magnetic quantum number $m$. In an external magnetic field, the $m$ quantum number becomes observable due to the Zeeman effect, which separates the (initially degenerate) $m$ states in energy. This observation of $m$ in the energy domain however leaves the underlying circular electron motion invisible. Here we present an ultrafast ionization experiment in which we induce a directional ring current in a ground state ion by optical tunneling – the mechanism, which impacts atoms and solids in the same way[1]. We probe this ring current in a time delayed second ionization step, showing how the escaping electron maps its bound state circular motion onto a detector. In rare gas atoms, such as e.g. argon, orbitals of positive and negative $m$ with their clockwise and anti-clockwise sense of rotation are equally populated and no net ring current remains. If one, however, finds a process that selectively ejects an electron from only one of these energetically degenerate orbitals, then the remaining ion will possess a stationary ring current with defined sign, even in its ground state. Similar ground-state currents emerging due to optical tunneling in condensed matter systems, e.g. in solid argon, could result in topological edge currents in a manner similar to their appearance in twisted waveguides[14], opening an exciting opportunity for ultrafast imaging and control of their formation in condensed matter systems.

Recent theoretical works predicted that optical tunneling through a rotating barrier depends on the sign of the magnetic quantum number $m$. Such a barrier can be created by a strong circularly polarized laser pulse impinging on the atom (Fig. 1). Counterintuitively, theory predicts that electrons, which are counter-rotating with respect to the tunneling barrier, are strongly preferred for tunnel ionization[4,6,15]. Using this insight, the generation of spin polarized electrons has been predicted[15] and measured recently[16]. Another experiment demonstrated that sequential double ionization rates by two subsequent circular laser pulses increase if their polarizations are counter-rotating, compared to the case when both rotate in the same direction[17]. In the present work we directly prove and quantify this quantum state selectivity of optical tunneling in a pump-probe experiment. To distinguish emitted electrons with different $m$ we have employed the following simple idea. Suppose an electron escaping from the tunnel into the continuum keeps its original angular momentum. Then at the position of the tunnel exit $r_t$ an angular momentum of $m\hbar$ corresponds classically to a linear momentum of $p_\perp = m\hbar/r_t$ perpendicular to the tunnel direction. Thus, the electron starts its motion in the field with an additional initial momentum. Depending on the sign of $m$ with respect to the sense of rotation of the laser field this initial momentum either adds to or subtracts from the laser-induced drift momentum. Measuring the momentum distribution should then give direct access to stationary ring currents present in a single atom as well as to the "transport" of angular momentum through the tunneling barrier.

In spite of the seeming simplicity of the idea, visualizing these ring currents in an experiment is rather demanding. Two circularly polarized laser pulses have to be employed. The pump laser pulse should generate ring current in the ion. To detect this current, we apply a probe laser pulse to remove the second electron that should carry the fingerprint of the current in its energy spectrum. A major challenge in the experiment is to identify the atoms which have been ionized subsequently by the pump and the probe pulse and not have been doubly ionized by either of the two. Accordingly, the properties of the pump and the probe pulses need to differ such that the measured electrons carry information on their ionization sequence. In our pump-probe experiment a circularly polarized pump pulse with a wavelength of $\lambda$ =390 nm ejects a first electron and a delayed more intense elliptically polarized probe pulse with a wavelength of $\lambda$ =780 nm ejects the second electron from an argon atom. We measure the changes in ionization rate and the momentum distribution of the second electron upon switching the helicity of the pump pulse, while the probe pulse remains the same in all cases. Figs. 1 and 2 show the pump and probe step in more detail. For $t = 0$ fs the circularly polarized ultrashort pump pulse with an intensity of $I_0 = 2.1 * 10^{14}$ W/cm$^2$ creates a singly charged argon ion. About 200 fs later, long after the pump pulse is gone, the second electron is emitted from the ion by the probe pulse (ellipticity of 0.61 with a peak electric field of $F_L = 0.11\ a.u.$, corresponding to an intensity of $I_L = 8.5 * 10^{14}$ W/cm$^2$ for circularly polarized light). Since the liberated electrons are accelerated by the laser field, their final momenta are proportional to $|F_L * \lambda|$ which is very different for both pulses. Thus, one can tell from the final momentum of each electron whether it was ejected in the pump or in the probe step. Moreover, the coincident detection of electrons and ions allows us to identify and reject events, where both electrons are set free by the same pulse, as described in more detail in the methods section.

Fig. 1 displays the momentum distributions of the electron emitted by the pump pulse with counter-clockwise rotating polarization (Fig. 1c, indicated by "L") and clockwise rotating polarization (Fig. 1f, indicated by "R"). As both $m$-states are equally populated in the neutral Ar atom, the momentum distributions and ionization rate are independent of the sign of the helicity of the ionizing laser field. However, the memory about the sense of rotation of the ejected electron is recorded in the ion (Figs. 1, 2a and 2d). Will the second ionization step read out this memory? This ability depends on the frequency of the laser field. In the adiabatic tunneling limit, the barrier does not rotate during tunneling. Hence, tunneling from the states with positive and negative $m$ will be identical and the information about the initial sense of the electron's rotation will be lost. The multiphoton ionization regime offers a more optimistic picture: one would expect that the electron escapes with the angular momentum $I_p/\omega + m\hbar$, where $I_p/\omega$ is the angular momentum associated with the absorption of the minimal number of photons required to overcome the ionization potential $I_p$. Hence, in the intermediate, so-called non-adiabatic tunneling regime[2,3], the difference in the angular momenta $\Delta l_z$ for the $\pm|m|$ electrons just after tunneling should be between zero and $2|m|\hbar$ (see Extended Data Figs. 1 and 2 for semi-classical calculations). Our theory, which uses the analytical R-matrix (ARM) method[5,10,18–20] (see Methods), predicts that $\Delta l_z \approx \hbar$ in the conditions of our experiment (Figs. 2b and 2e). At the tunnel exit $r_t$ this difference corresponds to different transverse velocities, $\Delta v_\perp = \Delta l_z/r_t$, which translate into the different final momenta at the detector. Fig. 2 (2b, 2e) shows how the initial $m$ is transported through the tunnel and mapped onto the final electron spectrum.

We can now investigate the momentum distribution of the second electron $P^{2nd\ elec}$ for two scenarios. The electric field of the probe pulse rotates clockwise in all cases shown here (indicated by "R"). Fig. 2c (2f) shows $P_{LR}^{2nd\ elec}$ ($P_{RR}^{2nd\ elec}$) for the case where the first electron has been removed by a counter-clockwise (clockwise) rotating field. While at first glance both distributions appear to be similar, detailed examination unveils slight differences, which become prominent in the corresponding angle-integrated kinetic energy spectra $Y^{2nd\ elec}$ of the 2nd electron shown in Fig. 3a (see Extended Data Fig. 3 for two-dimensional differential momentum spectra). The measured peak energies are 22.3 eV for $Y_{RR}^{2nd\ elec}$ and 21.2 eV for $Y_{LR}^{2nd\ elec}$. This disparity not only signifies the presence of the ring current in the Ar ion, but also allows us to detect its direction, directly observing the propensity rules of the optical tunneling: the electron counter-rotating to the laser field is preferred, dominating at low energies in the spectra (Figs. 3a and 3b). Experimental results are in excellent agreement with our TDSE simulations and ARM theory (Fig. 3a).

The relative yield $R(E) = Y_{LR}^{2nd\ elec}(E)/Y_{RR}^{2nd\ elec}(E)$ (Fig. 3b) is close to 2 for low energy electrons, equals 1 at 31.2 eV and drops below unity at high energies. Since there are two electrons for each $m$-state in a $p$-orbital, $R(E)$ must fulfill $0.5 \leq R(E) \leq 2$. The observation that the relative yield is close to 2 for the low energy electrons allows to conclude, that the pump pulse - depending on its helicity - perfectly selects either $m = +1$ or $m = -1$ (see Methods). Using this insight we are able to obtain the energy dependent yields $Y_{m=-1}^{2nd\ elec}$ and $Y_{m=+1}^{2nd\ elec}$ for $m = -1$ and $m = +1$ electrons, see Fig. 3c. Not only the absolute yields for $Y_{m=-1}^{2nd\ elec}(E)$ and $Y_{m=+1}^{2nd\ elec}(E)$ are different, but also the peaks are shifted in energy by 3.5 eV, in good agreement with analytical theory and excellent agreement with numerical TDSE simulations.

Our findings demonstrate that the radial shift of the final momentum is a fingerprint of the ring current induced in the ion by optical tunneling. We have traced this effect back to the $m$-dependent "initial" transverse momentum at the tunnel exit. The substantial change of yield and energy of the second electron released by the unmodified probe pulse upon inverting the helicity of the pump pulse experimentally proves that the singly charged ion stores information about the helicity of the pump pulse. It also quantifies the transport of angular momentum through the rotating tunneling barrier. Thus, the often neglected role of the sign of the magnetic quantum number plays a major role in strong field ionization. We expect that this is not restricted to atoms but will also influence molecular ionization.

The seminal paper of Keldysh[1] treated optical tunneling of electrons from atoms and multiphoton transitions between bands in solids on the same footing. Since then, the two fields, one concerned with intense laser-matter interaction in atoms, another in solids, developed separately. Fifty years later they have recognized each other's affinity[11,13,21–26]. The unification that followed has not only brought surprises, such as recognition of the crucial role of laser-driven electron recollisions in solids[23–25], but also immense opportunities for both fields[11,13,21–26]. Their remarkable affinity has allowed the extension of several key ideas of time-resolving the primary electronic response in atoms and molecules to condensed phase[23–26]. Extending these connections further towards inducing, time-resolving, and controlling such phenomena as topological states, the quantum Hall effect, and light-induced phase transitions is a dream. An atomic physics experiment described in this work opens an exciting opportunity for this dream.

# METHODS

**Laser Setup.** In order to generate the two laser pulses, we use a 200 µm $\beta$-barium borate crystal to double the frequency of a laser pulse with an initial wavelength of 780 nm (KMLabs Dragon, 40-fs FWHM, 8 kHz). A dielectric beam splitter separates the two pulses of different wavelengths. Subsequently intensity, polarization state and relative time delay are tuned by neutral density filters, λ/2 and λ/4 retardation plates and a delay stage, correspondingly. Both laser pulses are focused by a spherical mirror ($f = 80$ mm) into a gas target of argon atoms, which is produced by a supersonic expansion of argon gas into the vacuum through a tiny 30 µm nozzle. The gas target was collimated to less than 10 µm along the axial direction in the laser focus to reduce focal averaging. The peak intensities in the focus were calibrated comparing the measured drift momenta of Ar photoelectrons ionized by the circularly polarized pump pulse with a wavelength of 390 nm to our TDSE calculations. For the intensity calibration of the elliptically polarized probe pulse ($\lambda = 780$ nm) a helium target was used to avoid saturation of single ionization. The uncertainty of this calibration method is estimated to be 10%.

**Particle detection.** Upon ionization the fragments are guided by a homogeneous electric field (18.0 V/cm) and a homogeneous magnetic field (10.4 G) towards time and position sensitive detectors. The lengths of the electron and ion arms were 378 mm and 67.8 mm, respectively. The detectors consist of two Multi-Channel-Plates (MCP) in chevron configuration with a radius of 60 mm and 40 mm for the electron and the ion side, respectively. For both detectors the MCP-stack is followed by a three-layer delay line anode (HEX) with an angle of 60° between layers as manufactured by RoentDek[27]. In this configuration the three dimensional momentum of the first electron that hits the detector and one momentum component in the plane of polarization ($p_z$-direction along time-of-flight) of the ion are measured in coincidence (Cold Target Recoil Ion Momentum Spectroscopy (COLTRIMS))[28]. Employing momentum conservation the undetected electron's momentum component in the time-of-flight direction was calculated (see following paragraph). Laser, optics setup and particle detection are the same as used in[29].

**Distinguishing electrons emitted by the pump and the probe step in momentum space and background subtraction.** The undetected electron's momentum component in z-direction $p_{z\_calc}$ is inferred using momentum conservation. Electrons emitted by the pump pulse have lower momenta (Figs. 1c and 1f) in $p_z$-direction than those generated by the probe pulse (Figs. 2c and 2f). Imposing the condition $|p_{z\_calc}| < 0.5$ a.u. for the calculated electron, we make sure that the electron that has been detected originates from the ion that was successfully ionized by the pump pulse before. The measured electrons for this condition are seen in Fig. 2. Since the electron momentum distribution is close to the negative vector potential of the probe pulse we know that those electrons stem from ionization by the probe pulse. For the spectra shown in Figs. 2, 3 and Extended Data Fig. 3 we have subtracted 35 % of random coincidences.

**Obtain experimental spectra for ionization from $m = +1$ or $m = -1$ electrons.** The relative yield from $Y_{LR}^{2nd\ elec}$ and $Y_{RR}^{2nd\ elec}$ (Fig. 3b) reaches values close to 2 for low energy electrons. This allows us to invert our data and obtain pure photoelectron distributions corresponding to either $m = +1$ or $m = -1$ electrons, removed at the probe step. Let w$^+$ (w$^-$) be the probability of liberating the electron co-rotating (counter-rotating) with the pump pulse. Consider first the case when the pump and the probe pulses rotate in the same direction. Since there are two electrons for each $m$-state in a $p$-orbital, the amount of counter-rotating electrons available for the second step is $N^- = 1 + \frac{w^+}{w^+ + w^-} = 1 + a$. Here $a$ is the relative chance of removing the co-rotating electron at the pump step. The amount of co-rotating electrons available for the second ionization step is $N^+ = 1 + \frac{w^-}{w^+ + w^-} = 2 - a$. Therefore, the photoelectron signal generated by the probe pulse in the co-rotating setup is

$$Y_{RR}^{2nd\ elec}(E) = (1 + a) * Y_{m=-1}^{2nd\ elec}(E) + (2 - a) * Y_{m=+1}^{2nd\ elec}(E) \quad \text{with } a \in [0,1] \quad (1)$$

where $Y_{m=\pm 1}^{2nd\ elec}(E)$ describe photo-electron yields for the probe step. Similarly, for the counter-rotating setup:

$$Y_{LR}^{2nd\ elec}(E) = (2 - a) * Y_{m=-1}^{2nd\ elec}(E) + (1 + a) * Y_{m=+1}^{2nd\ elec}(E) \quad (2)$$

The parameter $a$ determines the purity of the state prepared by the pump pulse with $a = 0$ beeing equivalent to perfect selection of a given $m$-state by the pump pulse. The relative yield $R(E) = Y_{LR}^{2nd\ elec}(E)/Y_{RR}^{2nd\ elec}(E)$ imposes stringent limitation on the value of $a$. Since $a$ is energy-independent, we can find its value from the measured low-energy behavior of $R(E) \to 2$ (see Fig. 3b) corresponding to $a = 0$ (theory yields $a_{TH} \approx 0.12$) and in this case $0.5 \leq R(E) \leq 2$ (the limits are marked in Fig. 3b). Setting $a = 0$ in equations (1) and (2) we obtain the energy dependent yields $Y_{m=-1}^{2nd\ elec}$ and $Y_{m=+1}^{2nd\ elec}$ for $m = -1$ and $m = +1$ electrons, see Fig. 3c.

**Numerical TDSE simulations.** Our numerical simulations are based on solving the time-dependent Schrödinger equation (TDSE) first for the pump step and then for the probe step, assuming sequential ionization. In both cases single active electron moves in an effective potential. For the pump step, two different effective potentials have been used, one developed and verified in Refs. 30 and 31 $V_{1,Ar}(r) = -(1. + 5.4\ exp(-r) + 11.6\ exp(-3.682r))/r$, and another defined as $V_{2,Ar}(r) = -[1 + 17exp(-r^2/1.1364)] / \sqrt{r^2 + 0.997exp(-2r^2)}$, both adjusted to fit the ionization potential of Argon from the p-shell. The numerical method used was described in Refs. 30 and 31 and is identical to that employed by us in Refs. 10, 16 and 20. For the pump step, only energy-integrated relative ionization yield from the co-rotating and counter-rotating orbitals is needed. For the estimated pump intensity (carrier frequency of $\omega = 0.114$ a.u., field strength $F = 0.055$ a.u., the ratio $a = w^+/(w^+ + w^-)$ was found to be $a = 0.18$ and $a = 0.12$ for the first and the second potentials correspondingly, demonstrating predominant ionization from the orbital counter-rotating to the pump field. For the probe step, the results shown in Fig. 3 and Extended Data Fig. 3 used $V_{3,Ar^+}(r) = -2/r + U(r)$, with the short-range part $U(r) = -6.24\exp(-1.235 \cdot r)/r$ adjusted so that the binding energy of the first p-state is equal to the ionization potential of $Ar^+$ ($I_p = 1.0153$ a.u.), and the next excited s-state ($E = -0.3845$ a.u.) approximates the first excitation in the $Ar^+$ ion. The calculations were performed for the orbitals co- and counter-rotating with respect to the probe field. The spectra for the second electron, for the co-rotating and counter-rotating pump-probe setups, have been obtained using equations (1) and (2) with $a = 0.12$.

The pulses used to produce the TDSE results in Fig. 3 and Extended Data Fig. 3 had carrier frequency $\omega = 0.057$ a.u., a $\cos^4(\pi t/NT)$ intensity envelope, with a full duration of $N = 6$ optical cycles $T = 2\pi/\omega$ (base to base). The pulses have an ellipticity $\epsilon = 0.61$, with $F_z = \epsilon F_y$. We have performed calculations for six intensities with $0.09 \leq F_y \leq 0.115$ a.u. in steps of $\Delta F_y = 0.005$ a.u. The results of the simulations were averaged over the carrier envelope phase (CEP) of the pulse, incremented in steps of $\Delta CEP = 0.05\pi$. The results were also averaged over the focal volume intensity distribution, assuming Gaussian focus and the gas jet much thinner than the length of the focal spot. Same procedure was used for the analytical R-matrix (ARM) calculations described below. The ARM and the TDSE results presented in Figs. 2, 3 and Extended Data Fig. 3 use $F_y = 0.11$ a.u. as the peak field strength.

The discretization box used for the simulations had a radial box size of 1000 a.u., with $\Delta r = 0.1$ a.u. The maximum angular momenta included is $l_{max} = 160$, and the time step was $\Delta t = 0.036$ a.u. A complex boundary absorber was placed starting at 30 a.u. before the end of the simulation volume to avoid unwanted reflections from the boundaries. The convergence of the numerical calculations has been checked with respect to all discretization parameters. The photo electron spectra were calculated by propagating the wave function one extra cycle after the end of the pulse, applying a spatial mask with a radius of 75 a.u. to remove the bound part of the wave function. The remaining (continuum) part was then projected on the well-known exact continuum eigenstates of the doubly charged Coulomb center. The accuracy of this procedure has been monitored by varying the extra propagation time up to five cycles, and by varying the radius and the width of the spatial mask.

In order to compare with the experimental measurements, we have used an angular filter, in the same way as is in the experiment, imposing two 90 degrees integration windows, centered at 100 degrees and 180+100 degrees respectively.

**Analytical R-matrix theory.** The Analytical R-Matrix (ARM) approach has been described in detail in Refs. 5, 6, 10, 18-20, with Ref. 6 focusing on its application to strong-field ionization from orbitals with non-zero $l, m$. The ARM method yields the following expressions for the photo-electron signal at the momentum $\boldsymbol{p}$

$$|a(\boldsymbol{p})|^2 = |R_{lm}(\boldsymbol{p})|^2 e^{2\text{Im}S(\boldsymbol{p},t_s)} \quad (3)$$

The second term encodes the bulk of the 'weight' of the quantum trajectory defined by the initial coordinate (at the origin) and the final momentum $\mathbf{p}$ at the detector. The trajectory leaves the bound orbital at a complex-valued time $t_s = t_s(\mathbf{p})$ and moves according to the Newton equations, both in the classically forbidden and classically allowed regions. Extension into the classically forbidden region makes the starting time $t_s = t_s(\mathbf{p})$ complex-valued. The time $t_s = t_s(\mathbf{p})$ is found as the solution of $\partial S(\mathbf{p}, t)/\partial t = 0$, where the action $S(\mathbf{p}, t_s)$ is calculated along the complex-valued trajectory and is complex-valued. The strength of the photo-electron signal depends on its (negative) imaginary part $\text{Im} S(\mathbf{p}, t_s)$. The action includes the electron interaction with the laser field and the core potential. Further details are briefly summarized in equations (2-6) of the supplementary part of Ref. 10, with complete mathematical treatment presented in Refs. 6, 5, 18-20, including the verification against ab-initio simulations of the time-dependent Schrödinger equation (5).

The first term in Eq. (3) encodes the angular structure of the ionizing orbital, $R_{lm}(\mathbf{p}) \propto e^{im\varphi(\mathbf{p})}$, where $\varphi(\mathbf{p})$ is the complex-valued 'tunneling' angle – the angle at which the trajectory leaves the origin, $\tan \varphi(\mathbf{p}) = v_y(t_s(\mathbf{p}))/v_x(t_s(\mathbf{p}))$. The angle is complex-valued due to the complexity of the velocity in the classically forbidden region, and its imaginary part determines the relative ionization yields from orbitals with $m = \pm|m|$: $|R_{l,+|m|}(\mathbf{p})|^2/|R_{l,-|m|}(\mathbf{p})|^2 \propto e^{-4|m|\text{Im}\varphi(\mathbf{p})}$. The effects of the core potential on the outgoing electron are included in the action and in the shift of the ionization time $t_s = t_s(\mathbf{p})$. The corrections to the tunneling angle $\varphi(\mathbf{p})$ associated with the effect of the core potential were not included.

**Semiclassical calculation.** The semiclassical simulation of ionization for the argon ion by the 780 nm probe pulse is based on the semiclassical two-step (SCTS) model of Ref. 32. The initial conditions (ionization time and transverse momentum) for each trajectory are prepared using importance sampling[32] according to the Ammosov-Delone-Krainov ionization theory[33]. The tunnel exit is obtained in the same way as in[34]. A linear offset momentum $p_\perp = m\hbar/r_t$ in the plane of polarization and perpendicular to the tunneling direction is added to the momentum distribution. The momentum $p_\perp$ corresponds to an angular momentum of $m\hbar$ at the position of the tunnel exit $r_t$. The results from our ARM calculation on the transverse momentum distribution at the tunnel exit shown in Figs. 2b and 2e show a momentum difference of only 0.12 a.u. between the ionization of $m = +1$ and $m = -1$ which is about a factor of two smaller than the value we assume here based on the classical estimate $p_\perp = m\hbar/r_t$. After tunneling an electron is propagated classically in the presence of the doubly charged ionic core and the strong laser field. The analogue of the quantum mechanical phase was calculated from classical action for each final momentum. A peak electric field of $F_{y\_classical} = 0.114$ a.u. and an ellipticity of $\epsilon = 0.61$ ($F_{z\_classical} = \epsilon F_{y\_classical}$) have been used. The yield from the semiclassical simulations in Extended Data Figs. 1 and 2 have been normalized to fit the maximum of the experimental data for $m = -1$ and $m = +1$ respectively.

**Differences in angle-resolved photoelectron spectra.** The photoelectron spectra in Figs. 2c and 2f show - besides the discussed radial differences - also angular deviations. Those are due to the Coulomb attraction of the outgoing electron to the ionic core. The different "initial conditions" map onto different angular offsets in the final angle-resolved photoelectron spectra originating from $m = +1$ and $m = -1$ states[6], similar to the so called attoclock set-up[7,12,35,36]. In our experiment this directly translates (see equations (1) and (2) into different angular structures in $P_{LR}^{2nd\ elec}$ and $P_{RR}^{2nd\ elec}$ (see Extended Data Fig. 3 for the differential histograms). The additional angular offset associated with the angular momentum of the ionizing state directly affects the interpretation of attosecond measurements of tunneling dynamics via the attoclock set-up[36] and also extends attoclock-type measurements[7,12,35,36] to multi-cycle laser pulses.

Accordingly, our findings give a purely experimental answer to the lively debated question regarding the role of non-adiabatic electron dynamics during optical tunneling[7,10]. So far, these effects have only been addressed by comparing experimental observations with calculated ones, using theoretical models to reconstruct the underlying dynamics[7–9,12,35–38].

**Data Availability.** The data that support the findings of this study are available from the corresponding author upon reasonable request.

**ACKNOWLEDGEMENTS**

This work was supported by the DFG Priority Programme "Quantum Dynamics in Tailored Intense Fields."

**AUTHOR CONTRIBUTIONS**

S.E., M.K., M.R., A. H. J.R., F.T., K.F., N.S., K.H., L.S., T.J., M.S., and R.D. contributed to the experiment. S.E., M.K., M.R., K.L., I.B., J.K., F.M., M.I., O.S., and R.D. contributed to the theoretical results. S.E., M.K., T.J., M.S., and R.D. performed the analysis of the experimental data. All authors contributed to the manuscript.

**AUTHOR INFORMATION**

Reprints and permissions information is available at www.nature.com/reprints. The authors declare no competing financial interests. Correspondence and requests for materials should be addressed to S.E. (eckart@atom.uni-frankfurt.de) or R.D. (doerner@atom.uni-frankfurt.de).

**Figure 1**

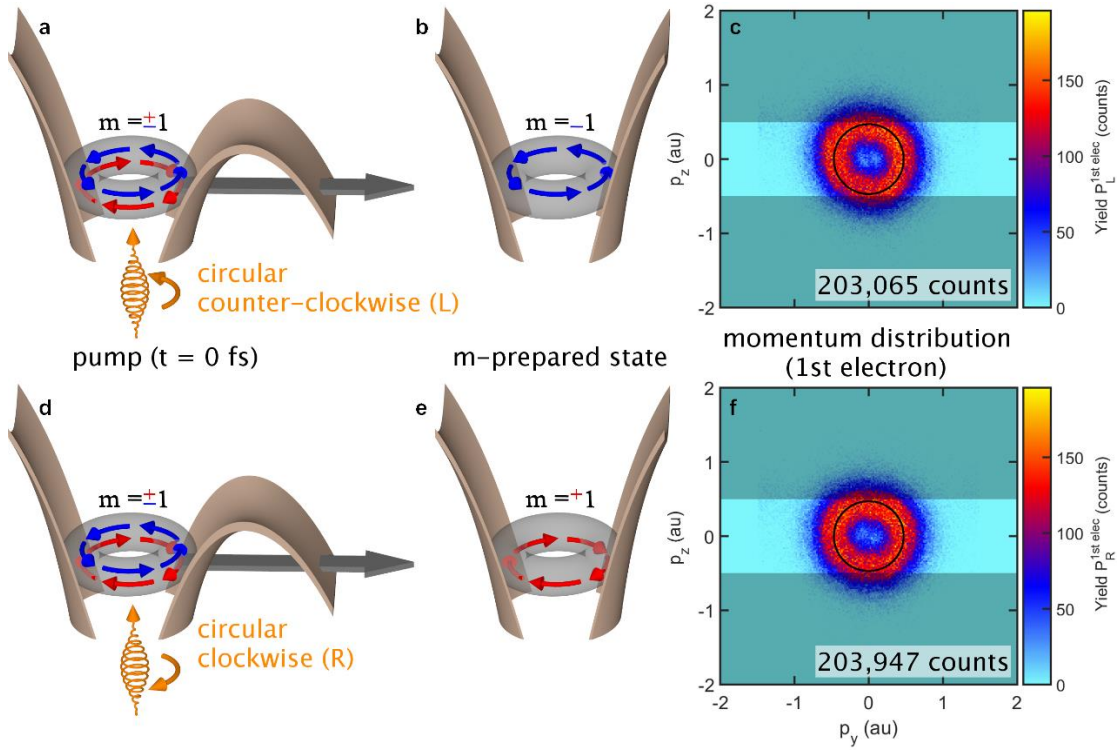

**Figure 1 | Experimental preparation of ring currents by $m$-dependent tunnel-ionization.** **a**, The circular polarized laser pulse with a counter-clockwise rotating electric field liberates an electron by tunnel-ionization at $t = 0$ fs. Clockwise rotating electrons ($m = +1$) are strongly preferred for ionization. **b**, Step **a** results in a persistent ring current in the remaining ion. **d** and **e**, The corresponding process for a circular clockwise rotating electric field. The sign of $m$ and direction of the ring current are inverted. **c** and **f,** The electron momentum distributions $P_L^{1st\ elec}$ and $P_R^{1st\ elec}$ of the electrons liberated by the pump pulse for equal acquisition times are identical. The black lines show the negative vector potential of the pump pulse. Most of the first electrons fulfill $|p_{z\_calc}| < 0.5$ a.u. (indicated by the gray shaded area), which is utilized for distinguishing electrons from the pump and the probe pulse.

**Figure 2**

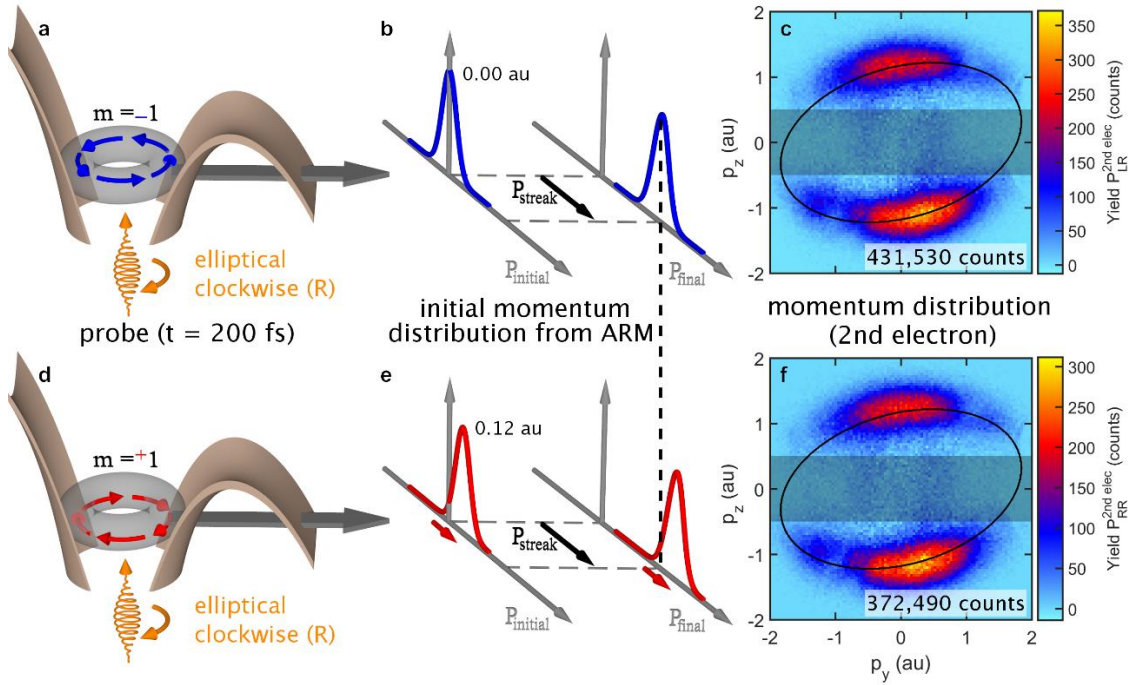

**Figure 2 | Detection of ring currents by momentum resolved $m$-dependent tunnel-ionization.**
**a** and **d**, At $t = 200$ fs the elliptically polarized probe pulse with clockwise rotating electric field hits the ion that has been created by the pump pulse (see Fig. 1). **c** and **f**, The second electron's momentum distributions $P_{LR}^{2nd\ elec}$ and $P_{RR}^{2nd\ elec}$ for equal acquisition times (the first electron, measured in coincidence, is selected to fulfill $|p_{z\_calc}| < 0.5$ a.u.). The momenta agree with the negative vector potential of the field (black line). There are more events in $P_{LR}^{2nd\ elec}$ than in $P_{RR}^{2nd\ elec}$, proving that the sign of the magnetic quantum number influences tunnel ionization. The angular and radial differences are discussed in Fig. 3 and Extended Data Fig. 3. The electron's "initial" momentum distributions after tunneling (dashed lines in panels **b**, **e**), with $m = +1$ has higher momentum (peak at $p = 0.12$ a.u.) than $m = -1$ (peak at $p = 0.00$ a.u.). These add to the drift momentum imparted by the laser pulse leading to different final radial momenta (solid lines in **b**, **e**). The transverse offsets for distributions in panels (**b**, **e**) are obtained using analytical R-matrix (ARM) theory, see text.

# Figure 3

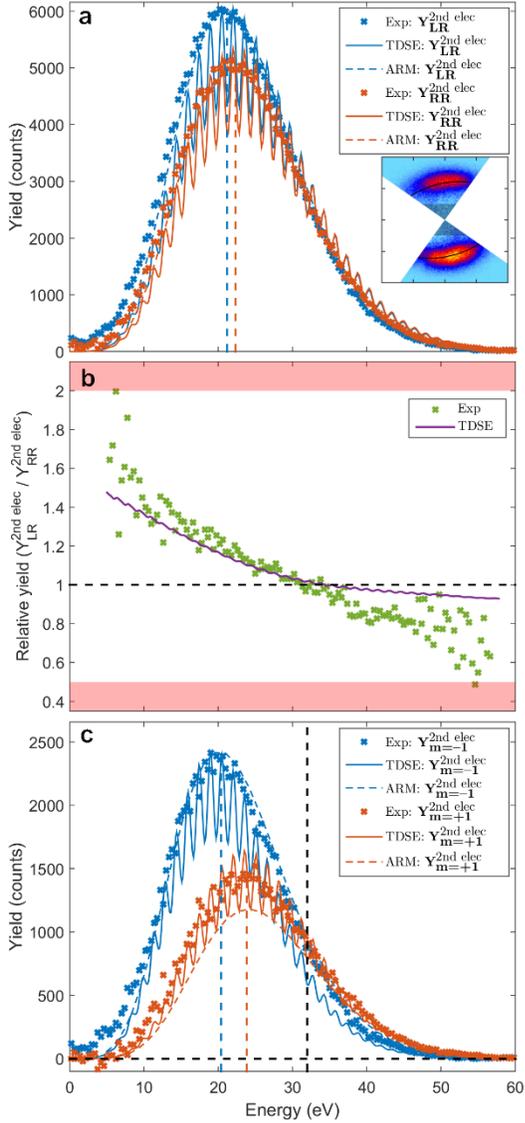

**Figure 3 | Energy resolved electron spectra showing ring current transport during tunneling.**
**a**, Measured (squares), theoretical (dashed lines) and numerically simulated (light solid lines) electron energy spectra for $Y_{LR}^{2nd\,elec}$ and $Y_{RR}^{2nd\,elec}$. To reduce noise, the momentum window shown in the inset is applied to the experimental data, the same window is used to gate theoretical and numerical results. **b**, The ratio of ionization rates $\frac{Y_{LR}^{2nd\,elec}}{Y_{RR}^{2nd\,elec}}$ is found to be between 0.5 and 2, the limits set by theory. Values close to 2 for low energy electrons indicate that the preparation by the pump pulse is almost perfectly selecting the sign of $m$. **c**, With the assumption of perfect preparation by the pump pulse, energy dependent yields from $Y_{m=-1}^{2nd\,elec}$ and $Y_{m=+1}^{2nd\,elec}$ according to equations (1) and (2) can be calculated directly from the measured data (squares) and are compared with our theoretical results (ARM theory, dashed lines) and numerical simulations of the time-dependent Schrödinger equation (solid lines), see supplementary material. Maxima are indicated by vertical colored lines.

**Extended Data Figure 1**

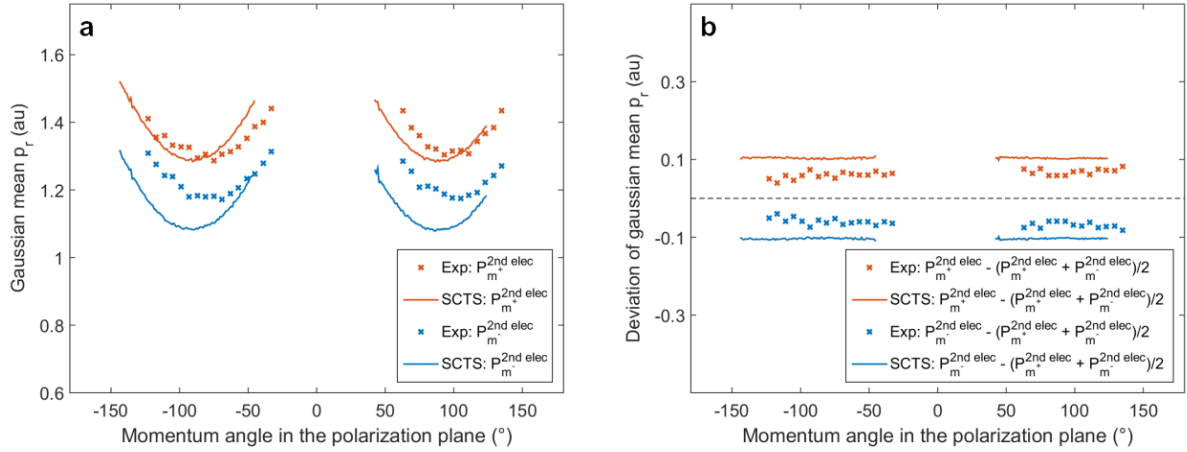

**Extended Data Figure 1 | Results of the semiclassical simulation. a**, Mean momentum (obtained by fitting a Gaussian distribution and only shown if experimental statistics are sufficient) vs. the momentum angle in the plane of polarization for the experimental data compared to the same quantity from the semiclassical simulations for $m = +1$ ($m^+$) and $m = -1$ ($m^-$). **b**, The same as **a** but as deviation to the mean of the $m^+$ and $m^-$. Besides a qualitatively good agreement it is found that the semiclassical simulation overestimates the momentum difference in comparison to the experimental results. This overestimation has its roots in higher differences in the initial transverse momenta for $m = +1$ and $m = -1$ states which are assumed in the semiclassical simulations.

**Extended Data Figure 2**

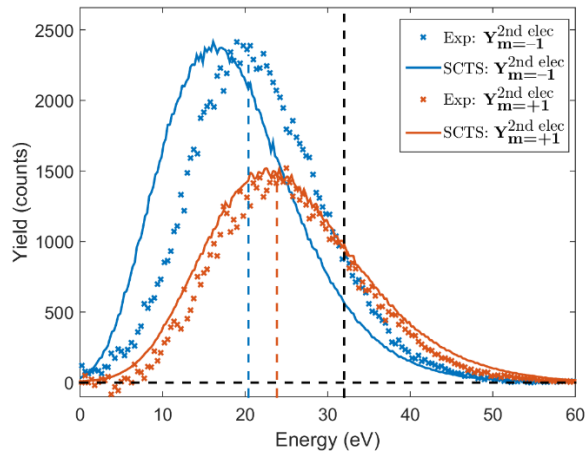

**Extended Data Figure 2 | Semiclassical simulation of the electron spectra.** Comparison of the experimental results from Fig. 3c with SCTS calculations. The semiclassical simulations assume that the tunneling electron preserves the magnetic quantum number $m$ during tunneling and hence has an initial transverse momentum $p_\perp = m\hbar/r$ at the tunnel exit. The peaks of the classical distributions are normalized to experimental data.

**Extended Data Figure 3**

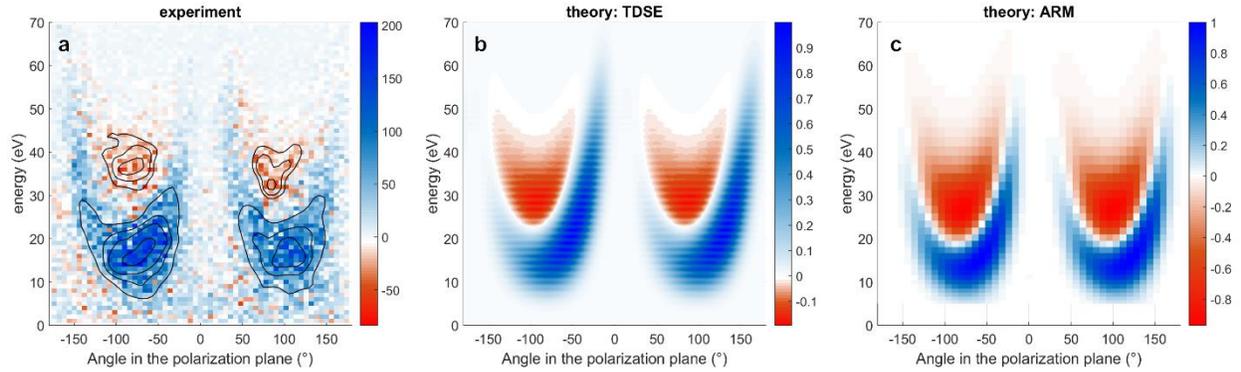

**Extended Data Figure 3 | Differences of the $m$-prepared state's electron momentum distributions.** Differential electron momentum distributions $P_{LR}^{2nd\ elec} - P_{RR}^{2nd\ elec}$ for experiment in **a**, numerical simulation of the time-dependent Schrödinger equation in **b** and analytical R-matrix theory (ARM) in **c**. The differences in the second electron's momenta (created by the probe pulse which is identical in all cases) are caused by the helicity of the pump pulse (which has created the ring current with defined sign in the singly charged ion). The radial differences divide the momentum distribution into a negative region and a positive region. The angular differences are due to different 'initial' momentum distributions of the electrons tunneling from states $m = -1$ (blue area) and $m = +1$ (red area). Black lines in **a** show contours to guide the eye.